\documentclass[twocolumn,showpacs,preprintnumbers,amsmath,amssymb,prl]{revtex4}

\usepackage{graphicx}
\usepackage{bm}

\begin{document}

\title{Inversion of Sequence of Diffusion and Density Anomalies in Core-Softened Systems}

\author{Yu. D. Fomin}
\affiliation{Institute for High Pressure Physics, Russian Academy
of Sciences, Troitsk 142190, Moscow Region, Russia}

\author{E. N. Tsiok}
\affiliation{Institute for High Pressure Physics, Russian Academy
of Sciences, Troitsk 142190, Moscow Region, Russia}

\author{V. N. Ryzhov}
\affiliation{Institute for High Pressure Physics, Russian Academy
of Sciences, Troitsk 142190, Moscow Region, Russia}

\date{\today}

\begin{abstract}
In this paper we present a simulation study of water-like
anomalies in core-softened system introduced in our previous
publications. We investigate the anomalous regions for a system
with the same functional form of the potential but with different
parameters and show that the order of the region of anomalous
diffusion and the region of density anomaly is inverted with
increasing the width of the repulsive shoulder.
\end{abstract}

 \maketitle

\section{I. Introduction}

It is well known that some liquids demonstrate anomalous behavior
in some regions of thermodynamic parameters. The most common and
well known example is water. The water phase diagrams have regions
where a thermal expansion coefficient is negative (density
anomaly), self-diffusivity increases upon compression (diffusion
anomaly), and the structural order of the system decreases with
increasing pressure (structural anomaly)
\cite{ad1,stanley1,deben2001,netz,book,book1}. Later on it was
discovered that many other substances also demonstrate similar
behavior. Some typical examples are silica, silicon, phosphorus
and many others
\cite{book,book1,ga,bi,te,s1,s2,be,ge,si1,si2,si3,s3,br,p1,p2}.

It is reasonable to relate this kind of behavior to the
orientational anisotropy of the potentials, however, a number of
studies demonstrate water-like anomalies in fluids that interact
through spherically symmetric core-softening potentials with two
length scales (see, for example, the reviews \cite{fr1} and
\cite{buld2009}). A lot of different core-softened potentials were
introduced.  All such systems can be approximately divided into
two classes: shoulder-like, composed of a hard-core and a
repulsive shoulder, softening the core, and ramp-like systems,
composed of a hard-core and a repulsive ramp, establishing two
competing equilibrium distances \cite{fr1,buld2009}. Anomalous
behavior usually takes place in ramp-like systems, while it is not
observed in purely shoulder-like ones. However, if instead of
"hard" step one considers a smoothed one the anomalies appear
(see, for example,
\cite{fr1,buld2009,fr2,stanley2,wejcp,wejcp1,wepre,wepre1,wepla}.

As it was found in experiments \cite{ad1} and simulations
\cite{deben2001,netz}, the water anomalies have a well-definite
sequence: the regions where these anomalies take place form nested
domains in the density-temperature \cite{deben2001} (or
pressure-temperature \cite{netz}) planes: the density anomaly
region is inside the diffusion anomaly domain, and both of these
anomalous regions are inside a broader structurally anomalous
region. This water-like behavior was found in systems with
core-softening potentials
\cite{fr1,buld2009,fr2,fr3,fr4,errington,errington2,st_bio,
8,9,10,11,12,13,14,15,16,17,18,19,20,bar2,21,22,23,24,25,26,
barb2008-1,yan2005,poole,PNAS1,PNAS2,widom1,barb2010,wejcp,wejcp1,wepre,wepre1,wepla}.
However, in other anomalous systems the sequence of anomalies may
be different. For example, the hierarchy of anomalies for silica
is different compared to water. In this case the diffusion anomaly
region contains the structural anomalous region which, in turn,
has the density anomaly region inside \cite{sio2}. To our
knowledge this is the only example of such inversion of the order
of the anomalies.

This paper presents a simulation study of anomalies in
core-softened system introduced in our previous publications
\cite{wejcp,wejcp1,wepre,wepre1,wepla}. We investigate the
anomalous regions for a system with the same functional form of
the potential but with different parameters and show that the
order of the region of anomalous diffusion and the region of
density anomaly is inverted.

The article is organized as following: Section II presents the
system and methods, Section III describes the results and gives
theirs discussion and Section IV contains conclusions.

\section{II. Systems and methods}

The system we study in the present simulations is Smooth Repulsive
Shoulder System (SRSS) introduced in our previous publications
\cite{wejcp,wejcp1,wepre}:

\begin{equation}
U(r)=
\varepsilon\left(\frac{\sigma}{r}\right)^{n}+\frac{1}{2}\varepsilon\left(1-\tanh\left(k_0
\left(r-\sigma_1 \right)\right)\right), \label{2}
\end{equation}
where $n=14,k_0=10$. $\sigma$ is "hard"-core diameter and
$\sigma_1=1.35;1.45;1.55;1.8$ is soft-core diameter.

In the remainder of this paper we use the dimensionless
quantities: $\tilde{{\bf r}}\equiv {\bf r}/d$, $\tilde{P}\equiv P
d^{3}/\varepsilon ,$ $\tilde{V}\equiv V/N d^{3}\equiv
1/\tilde{\rho}$. As we will only use these reduced variables, we
omit the tildes.

In Refs. \cite{wejcp,wepre} it was shown that this system
demonstrates anomalous behavior. A relation between phase diagram
and anomalous regions was also discussed in these articles. Our
later publications gave detailed study of diffusion, density and
structural anomalies in this system \cite{wepre1,wepla,we_cm}.

We simulated the systems with four different step sizes $\sigma_1$
and monitored the change in phase diagram and anomalous regions
with increasing the step width.

Importantly, in low temperature region the systems can demonstrate
slow dynamics. In order to get reliable results in the whole
temperature range we used parallel tempering technic
\cite{book_fs}. The system with $\sigma_1=1.35$ and
$\sigma_1=1.45$ were simulated at densities from $\rho=0.3$ till
$\rho=0.8$ with step $\delta \rho =0.05$ ($\delta \rho =0.025$ in
the vicinity of anomalous regions) and at temperatures from
$T=0.12$ till $T=0.8$. In the case of $\sigma_1=1.55$ the range of
densities was $\rho=0.2-0.75$. Finally, the density range of
$\rho=0.1$ - $0.8$ was simulated for $\sigma_1=1.8$. Along every
isochor we ran several parallel tempering runs with different
temperatures. Lowe-Andersen thermostat was used during the
equilibration time \cite{loweand}. After equilibration the
thermostat was switched off and the system evolved in $NVE$
ensemble for production. After the production a trial change of
temperature was made. Many data points were collected along every
isochor (more then $100$). Using these data points we constructed
$9-$th order polynomial approximation of internal energy, pressure
and diffusion coefficient along isochors. These approximations
were used in the following analysis.

In order to compute the excess entropy of the liquids we used
thermodynamic integration method \cite{book_fs}. We computed free
energy along a high temperature isotherm by integrating equation
of state of the system. After that we computed the free energies
along isochors by integrating the $U/T^2$ function, where $U$ is
the internal energy \cite{book_fs}. Excess entropy was obtained as
$S_{ex}=\frac{U-F_{ex}}{N k_BT}$.

\section{III. Results}

Here we present the anomalous regions for the four systems
studied.

\subsection {$\sigma_1=1.35$}

The phase diagram and anomalies in this system were already
reported in our previous publications
\cite{wejcp,wejcp1,wepre,wepre1}. Here we summarize the previous
results and place the anomalous regions on the phase diagram. We
also think that it is necessary to repeat the results for this
particular system for the sake of completeness.

\begin{figure}
\includegraphics[width=8cm, height=8cm]{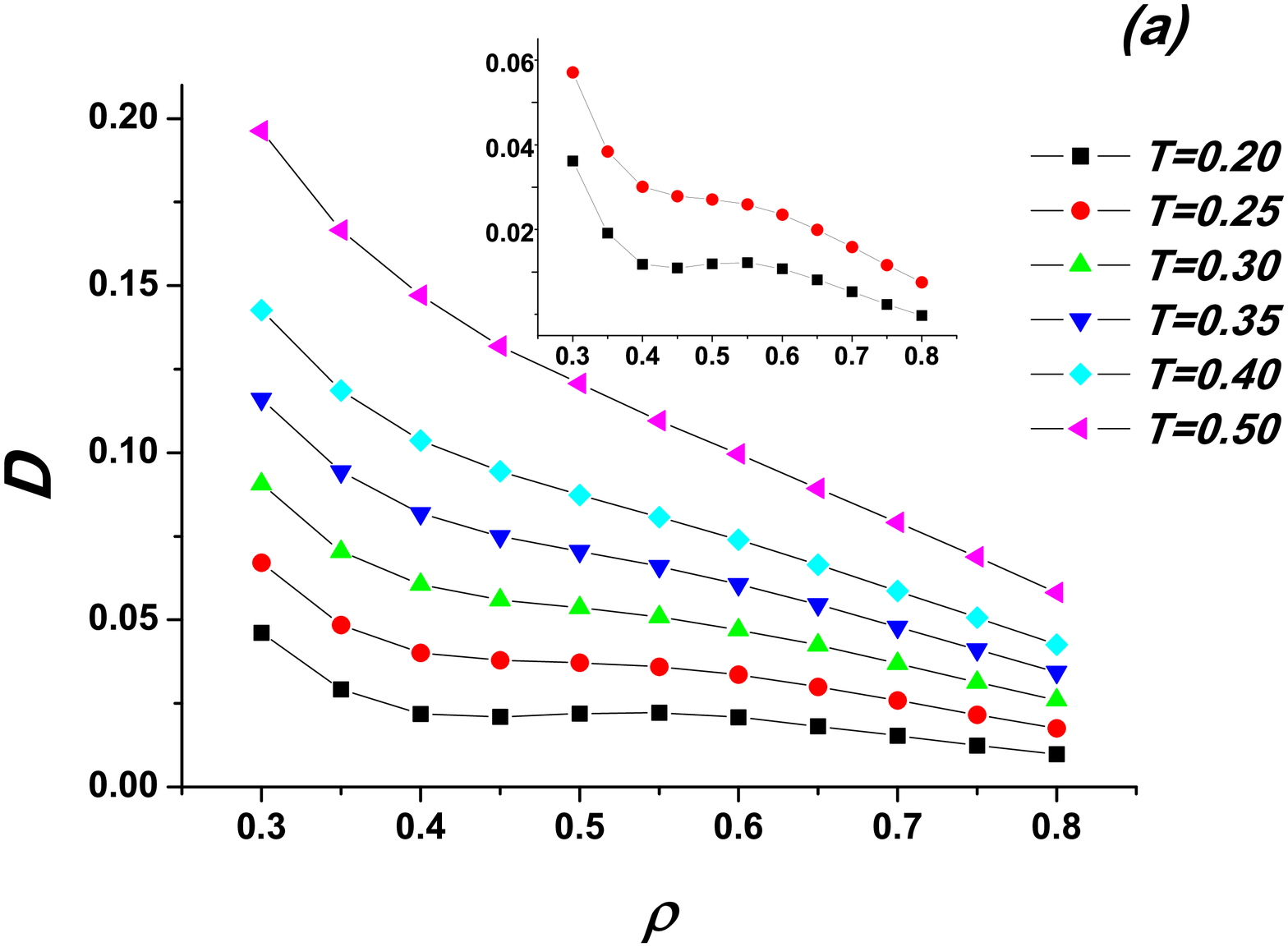}%

\includegraphics[width=8cm, height=8cm]{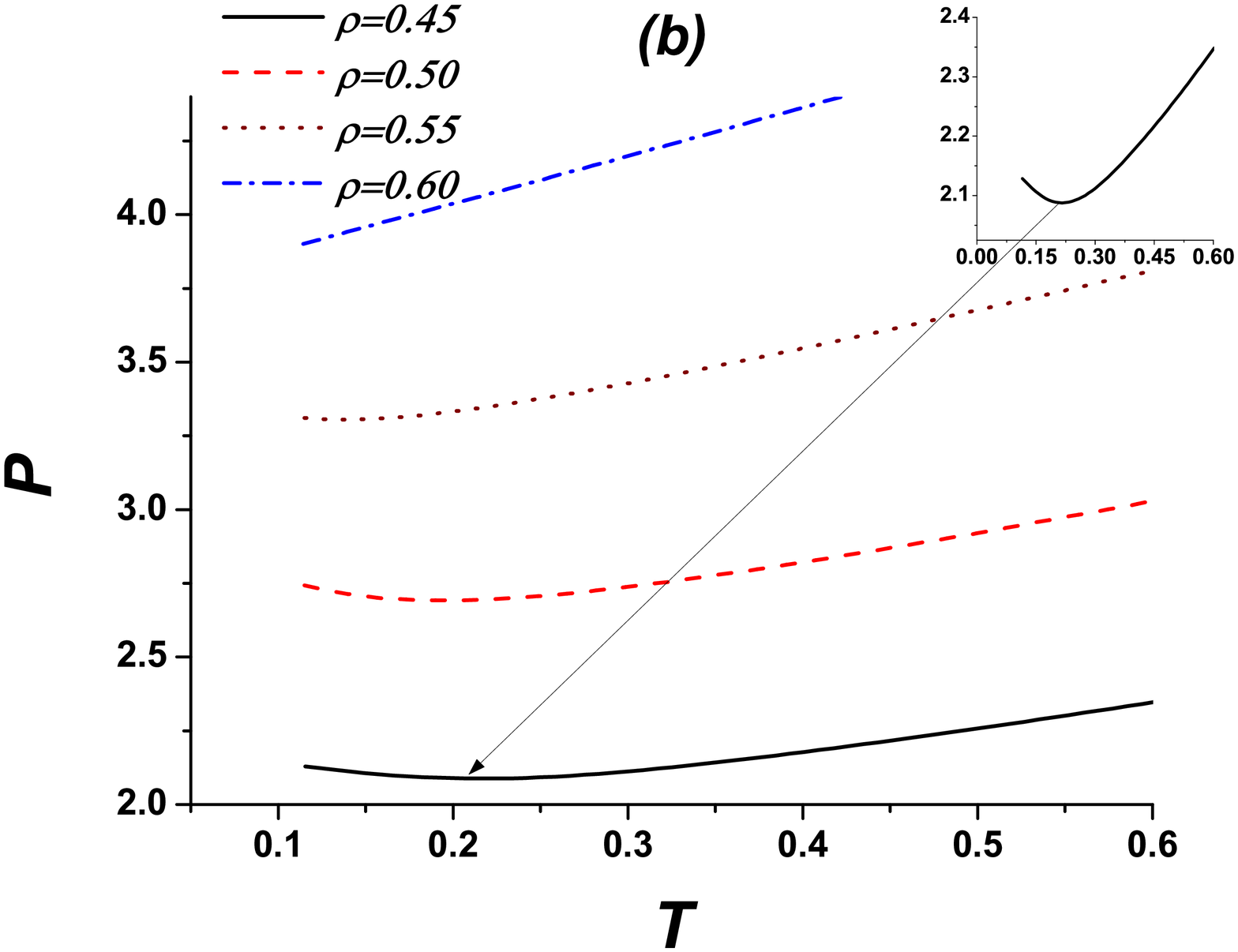}%

\includegraphics[width=8cm, height=8cm]{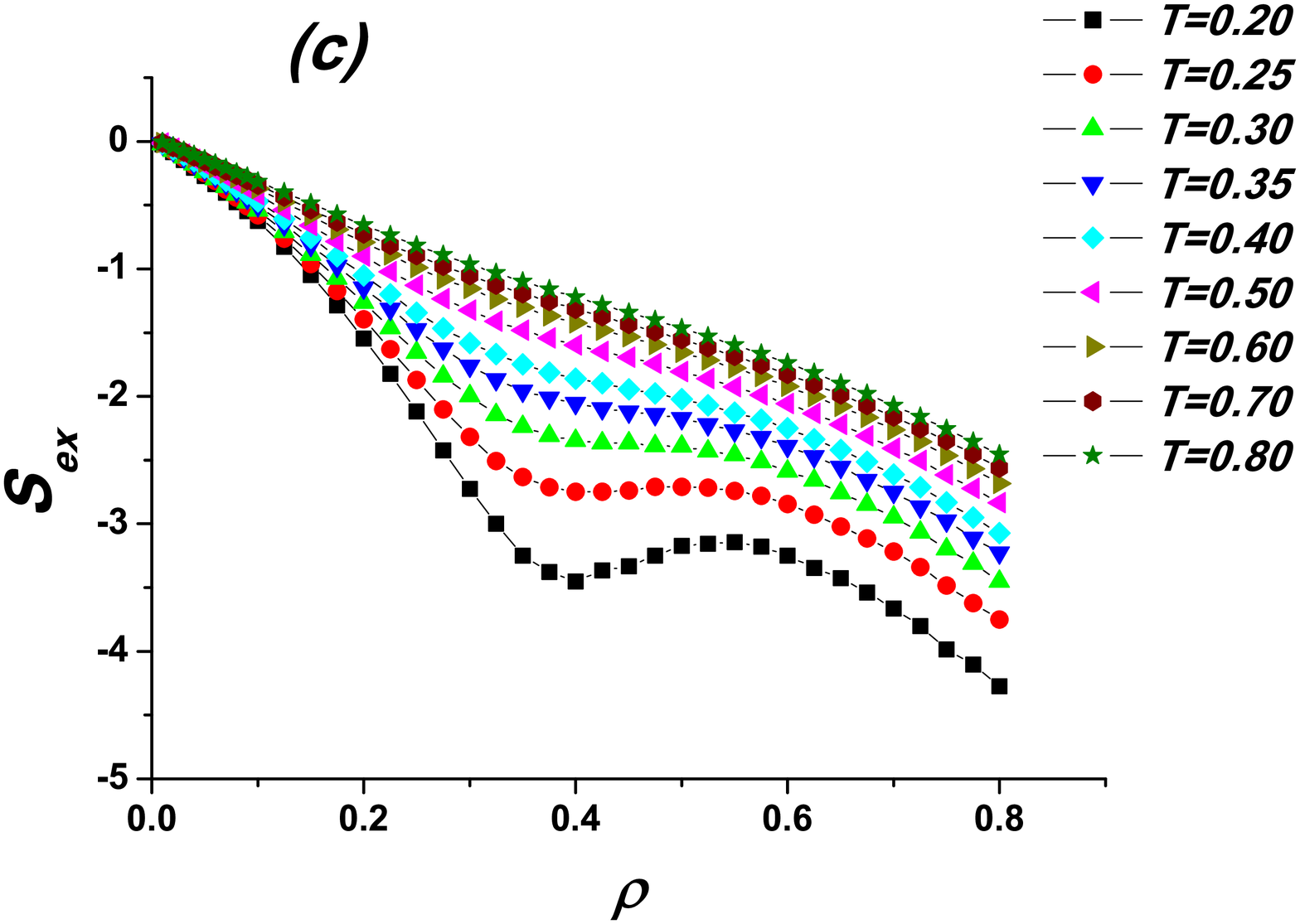}%

\caption{\label{fig:fig1} (Color online). (a) Diffusion
coefficient along a set of isotherms; (b) pressure along a set of
isochors; (c) excess entropy along a set of isotherms for the
system with $\sigma_1=1.35$.}
\end{figure}

Figs.~\ref{fig:fig1} (a)-(c) show the diffusion coefficient,
pressure and excess entropy for the system with $\sigma_1=1.35$.
One can see that all three anomalies take place in the system. It
is also evident that structural anomaly is more stable than the
diffusion one since it disappears at higher temperatures.

\begin{figure}
\includegraphics[width=8cm, height=8cm]{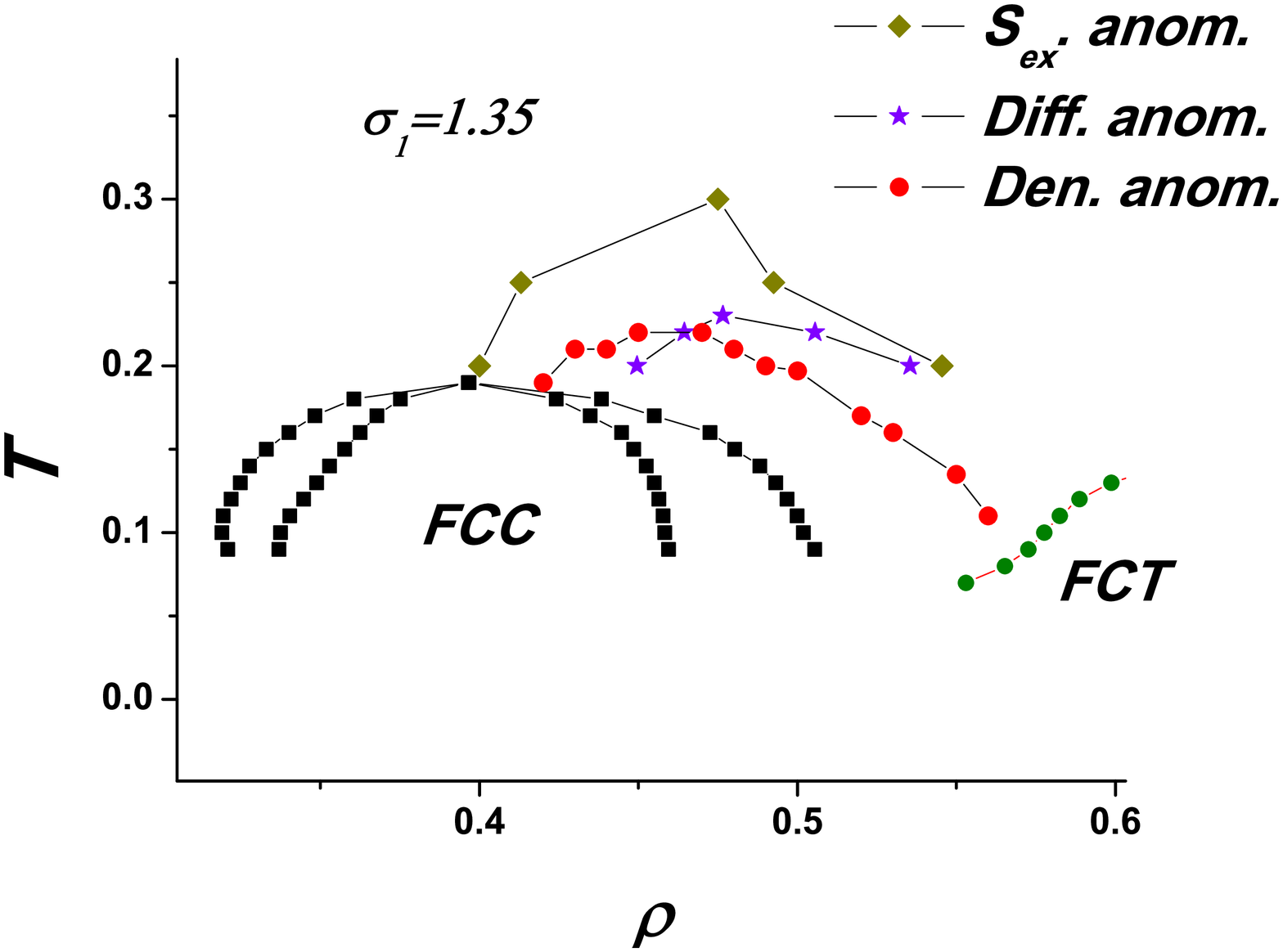}%

\caption{\label{fig:fig2} (Color online). Location of anomalous
regions at the phase diagram of the system with $\sigma_1=1.35$.}
\end{figure}

Fig.~\ref{fig:fig2} places the anomalies regions on the phase
diagram of the system reported in Refs. \cite{wejcp,wejcp1}. One
can see that all three anomalous regions appear exactly after the
low-density crystal bump. This allows to relate the anomalies to
the phase diagram, i.e. the reason for the anomalies to appear can
be the presence of the crystal bump inside the liquid phase. The
principle question here is that liquids to the left from the bump
and to the right have different local structures: while the
highest radial distribution function peak at the "left" liquid is
the second one, corresponding to the soft core distance, at the
"right" liquid it is the first peak located on the hard core
distance. In between of these two liquids a region appears where
strong competition of two distances appears. It is this
competition which brings to the appearance of anomalous behavior.

The second point to notice is that the anomalous regions
correspond to the picture proposed for water \cite{deben2001},
i.e. the diffusion anomaly region is inside the structural anomaly
and the density anomaly is mainly inside the diffusion anomaly.

\subsection {$\sigma_1=1.45$}

\begin{figure}
\includegraphics[width=8cm, height=8cm]{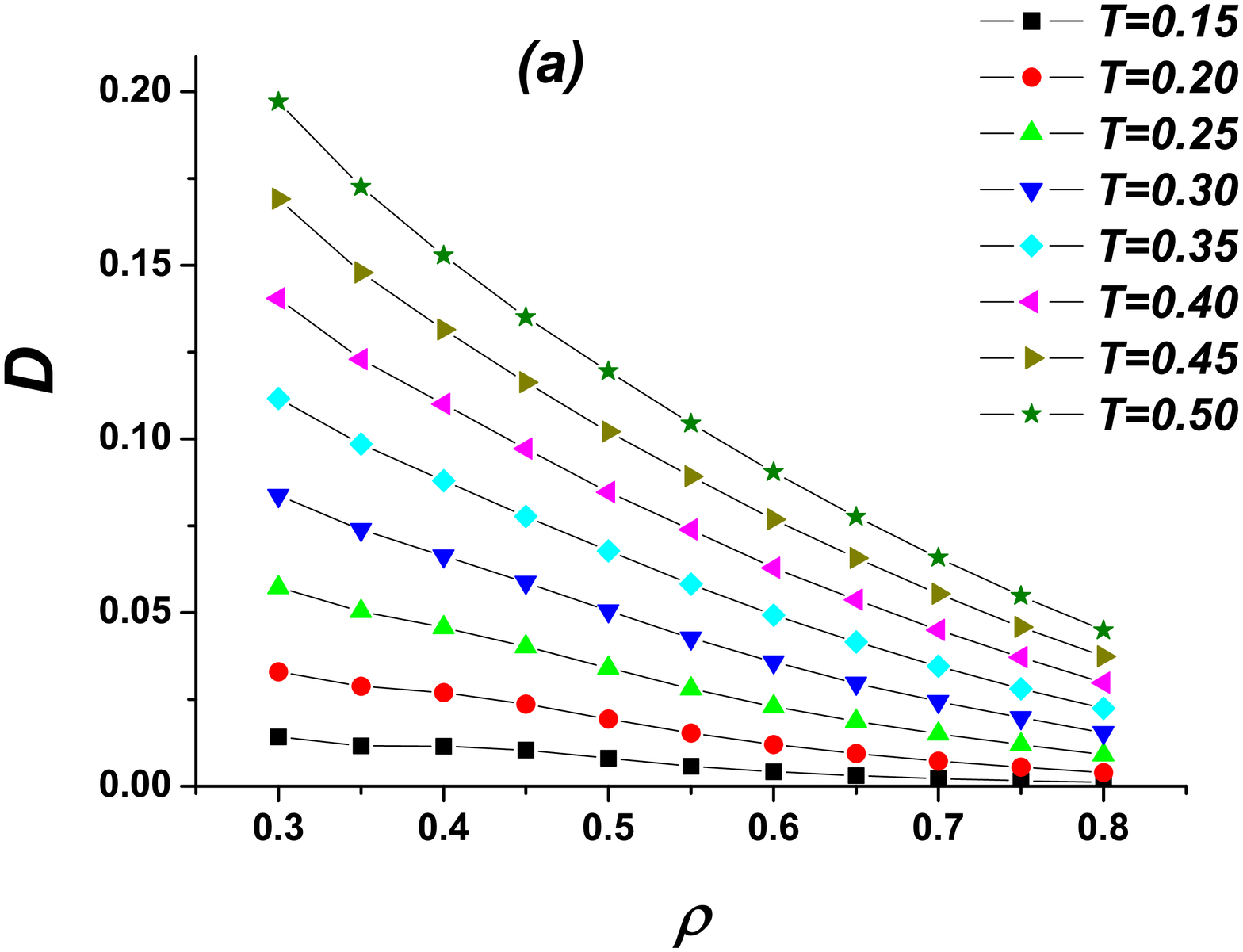}%

\includegraphics[width=8cm, height=8cm]{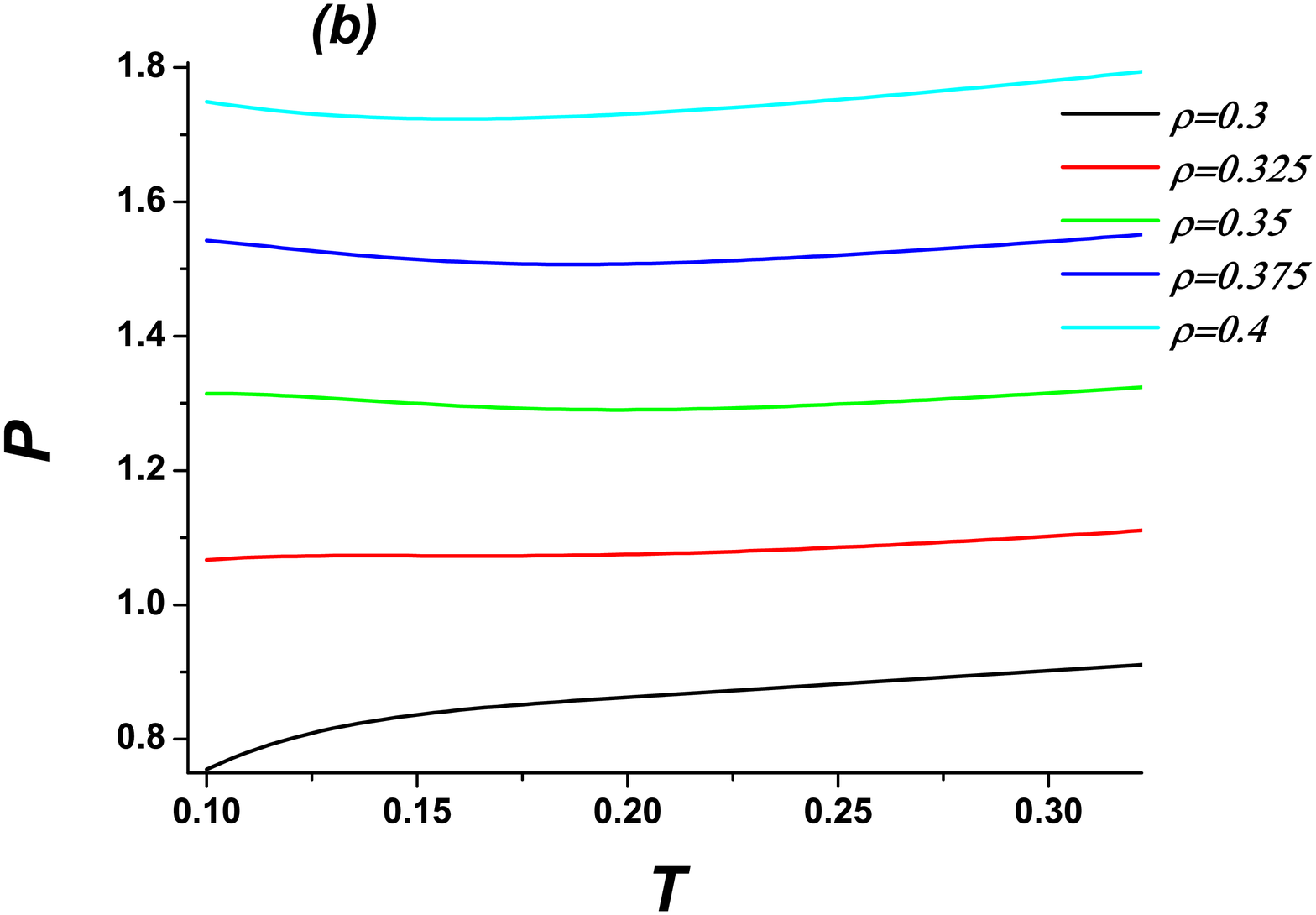}%

\includegraphics[width=8cm, height=8cm]{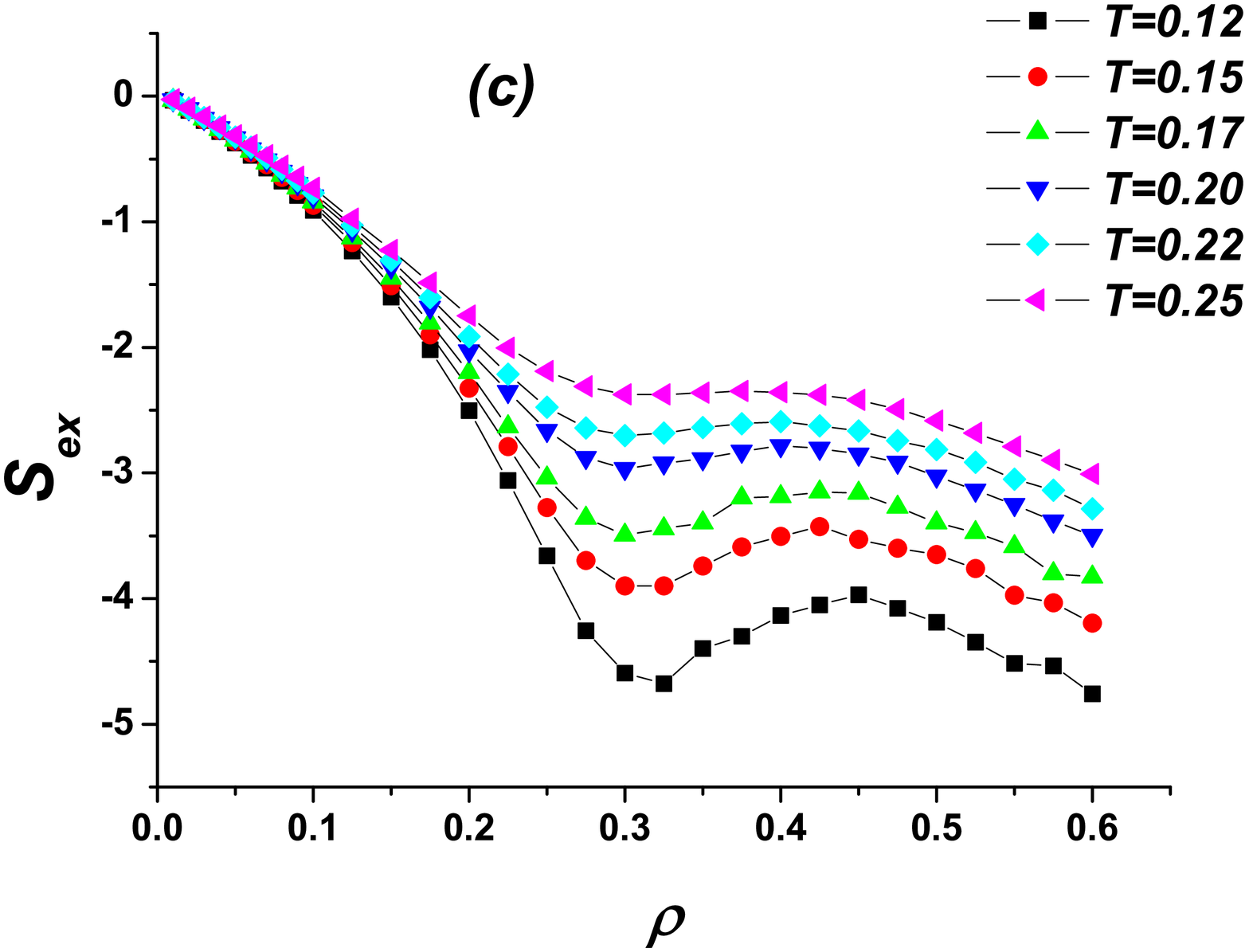}%

\caption{\label{fig:fig3} (Color online). (a) Diffusion
coefficient along a set of isotherms; (b) pressure along a set of
isochors; (c) excess entropy along a set of isotherms for the
system with $\sigma_1=1.45$.}
\end{figure}

Figs.~\ref{fig:fig3} (a)-(c) show the anomalies for the system
with $\sigma_1=1.45$. One can see from these pictures that the
system behaves qualitatively similar to the previous case.
However, the diffusion anomaly looks suppressed. At the
temperatures as low as $0.15$ the diffusion is close to a bend,
but it is still monotonous. At the same time the density anomaly
is rather pronounced in the system.

\begin{figure}
\includegraphics[width=8cm, height=8cm]{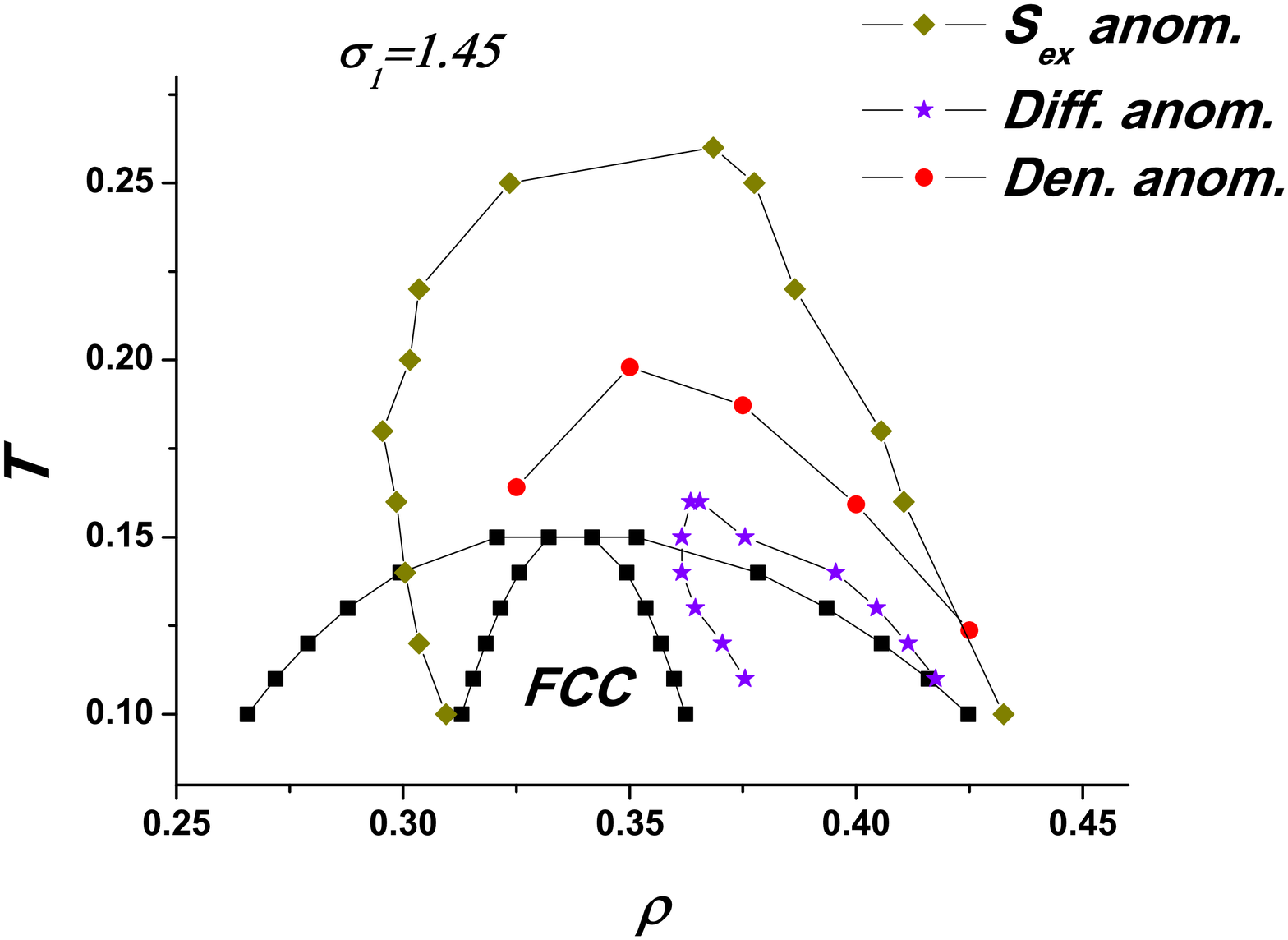}%

\caption{\label{fig:fig4} (Color online). Location of anomalous
regions at the phase diagram of the system with $\sigma_1=1.45$.}
\end{figure}

The location of anomalous regions and the low density Face
Centered Cubic (FCC) phase for $\sigma_1=1.45$ system is shown in
Fig.~\ref{fig:fig4}. One can see that the diffusion anomaly has
almost gone under the melting line. Only a small part of diffusion
anomaly region is located in stable liquid phase. At the same time
density anomaly is still very pronounced. It occupies a large
region in the liquid part of the phase diagram.

The most important fact extracted from Fig.~\ref{fig:fig4} is that
the diffusion and density anomalies inverted with respect to each
other, i.e. now diffusion anomaly region is inside the density
anomaly one. The similar inversion of the order of anomalies is
found in silica, however, in silica one can see the inversion of
the order of structural and diffusion anomaly \cite{si2}. We can
conclude that the case of unusual sequence of anomalies in silica
is not unique, and the SRSS system is one more example of such
unusual behavior.

\subsection {$\sigma_1=1.55$}

\begin{figure}
\includegraphics[width=8cm, height=8cm]{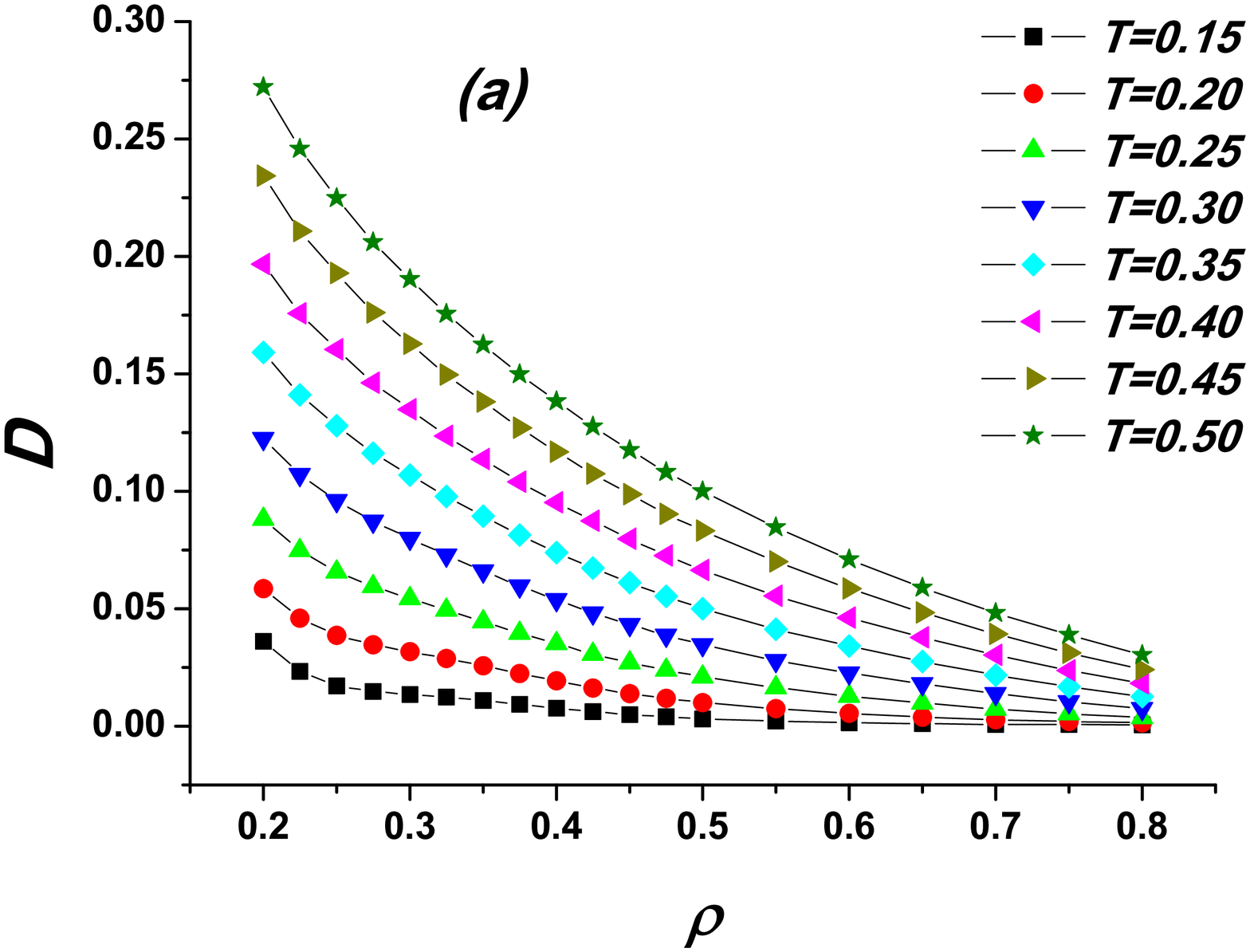}%

\includegraphics[width=8cm, height=8cm]{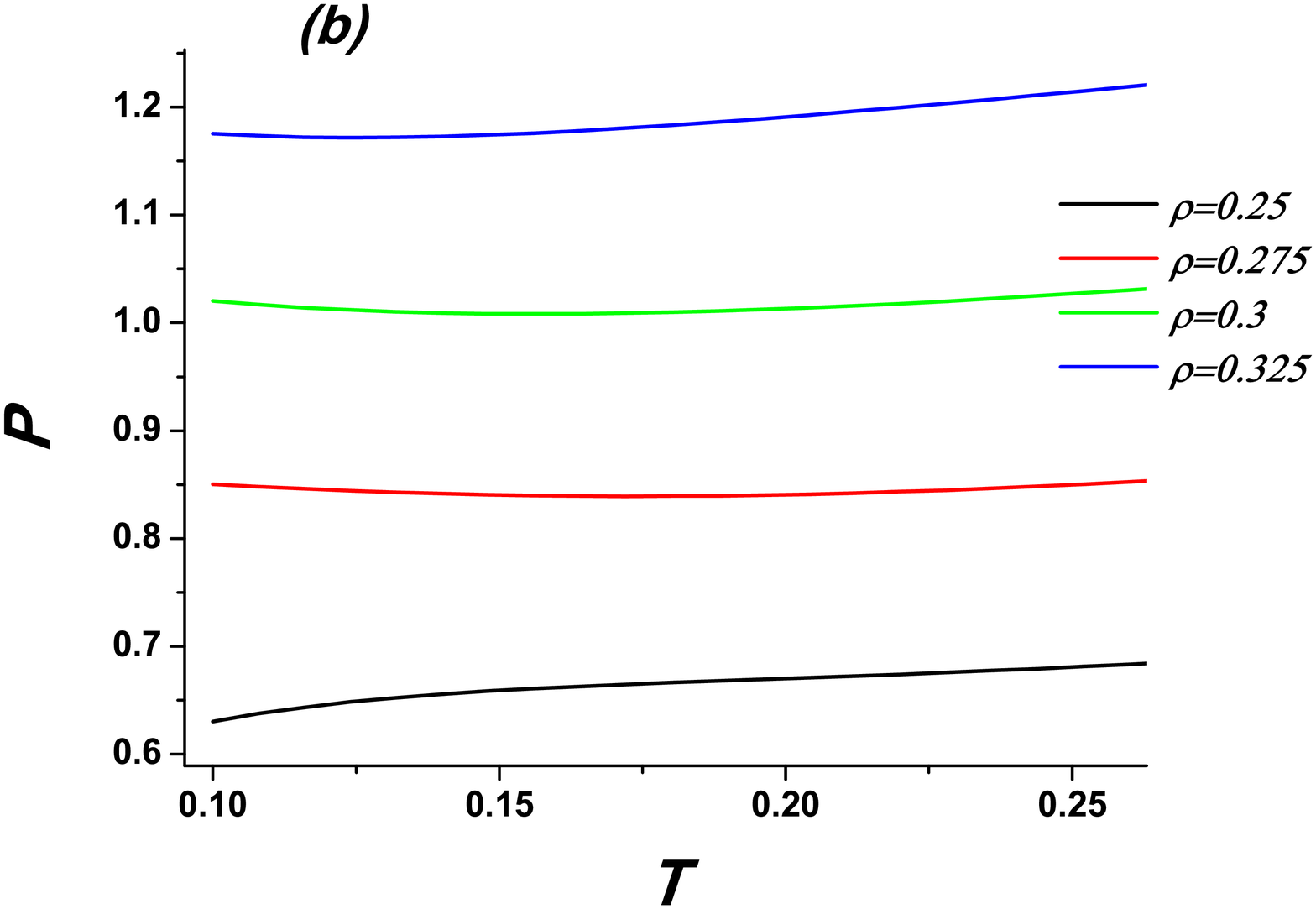}%

\includegraphics[width=8cm, height=8cm]{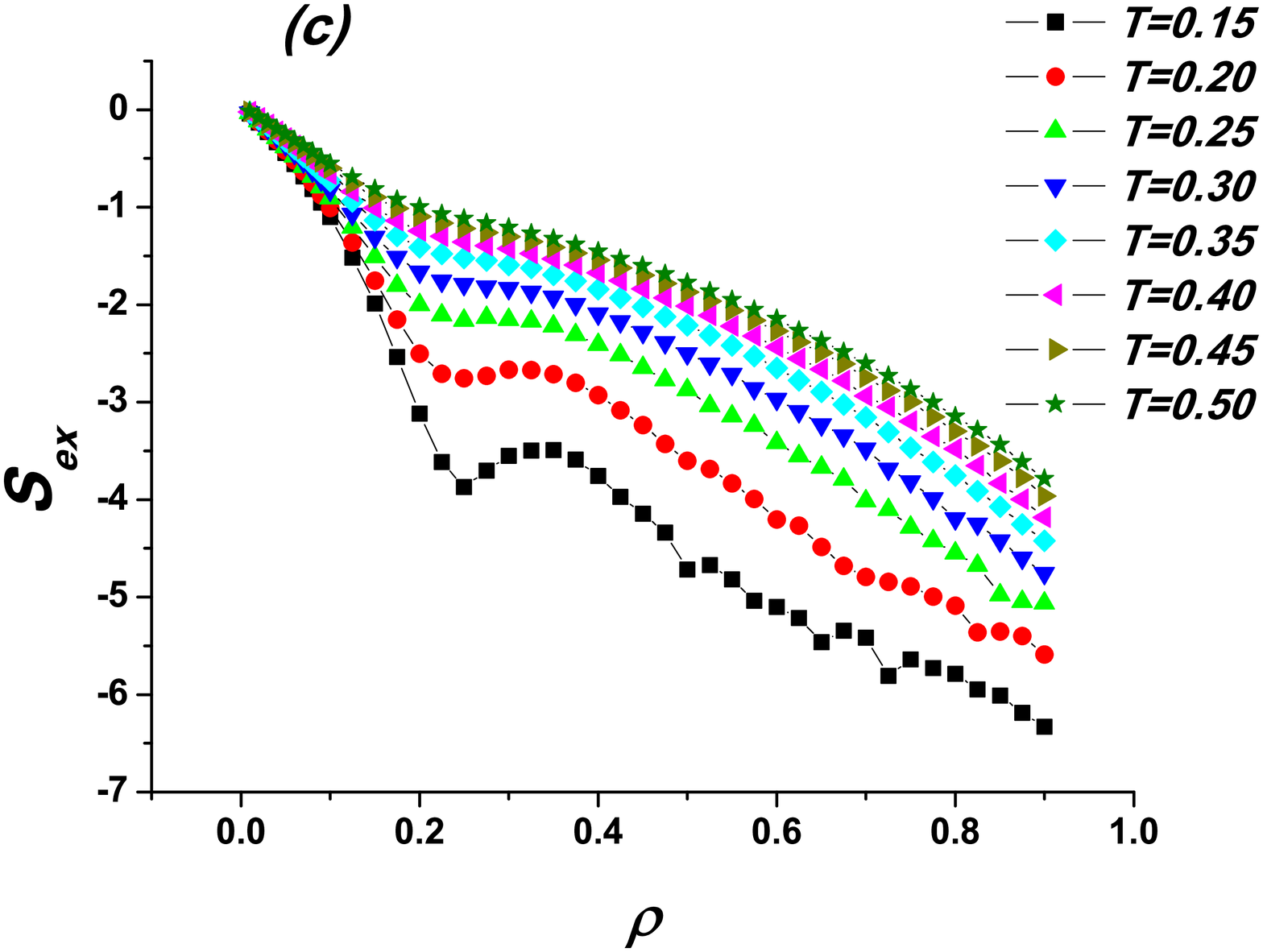}%

\caption{\label{fig:fig5} (Color online). (a) Diffusion
coefficient along a set of isotherms; (b) pressure along a set of
isochors; (c) excess entropy along a set of isotherms for the
system with $\sigma_1=1.55$.}
\end{figure}

In the case of $\sigma_1=1.55$ we do not find diffusion anomaly
(Fig.~\ref{fig:fig5}(a)). At low temperatures the diffusion along
an isotherm develops a bend, but not a loop. At the same time both
density and structural anomalies are present in the system
(Figs.~\ref{fig:fig5}(b)-(c)). Comparing it with the results for
the previous two cases we conclude that the diffusion anomaly is
completely under the melting line for the present case which makes
it unobservable.

\begin{figure}
\includegraphics[width=8cm, height=8cm]{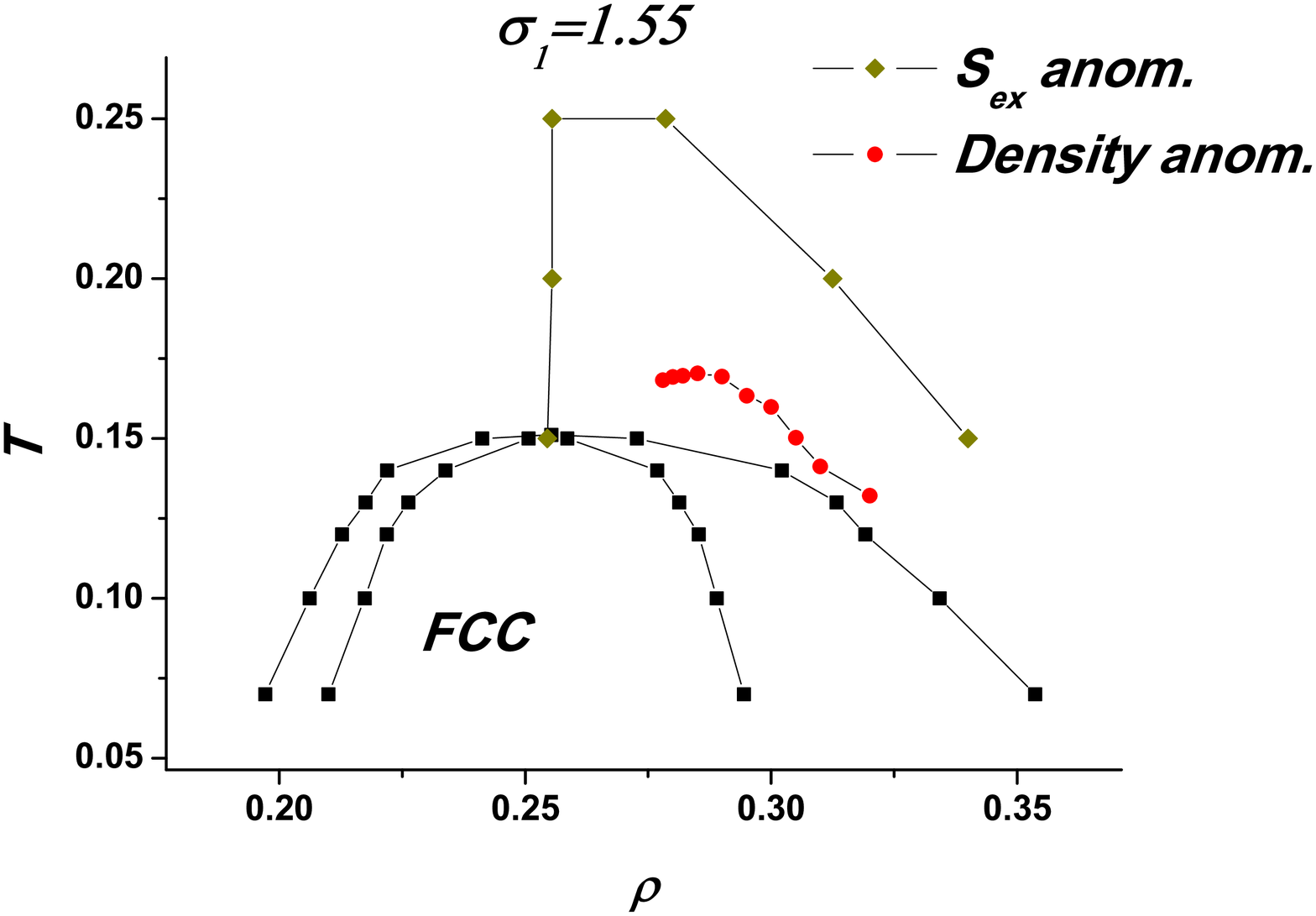}%

\caption{\label{fig:fig6} (Color online). Location of anomalous
regions at the phase diagram of the system with $\sigma_1=1.55$.}
\end{figure}

From Fig.~\ref{fig:fig6} we see that the density anomaly also
approaches the melting line. This allows to suggest that density
anomaly also shrinks with increasing the step size.

\subsection {$\sigma_1=1.8$}

\begin{figure}

\includegraphics[width=8cm, height=8cm]{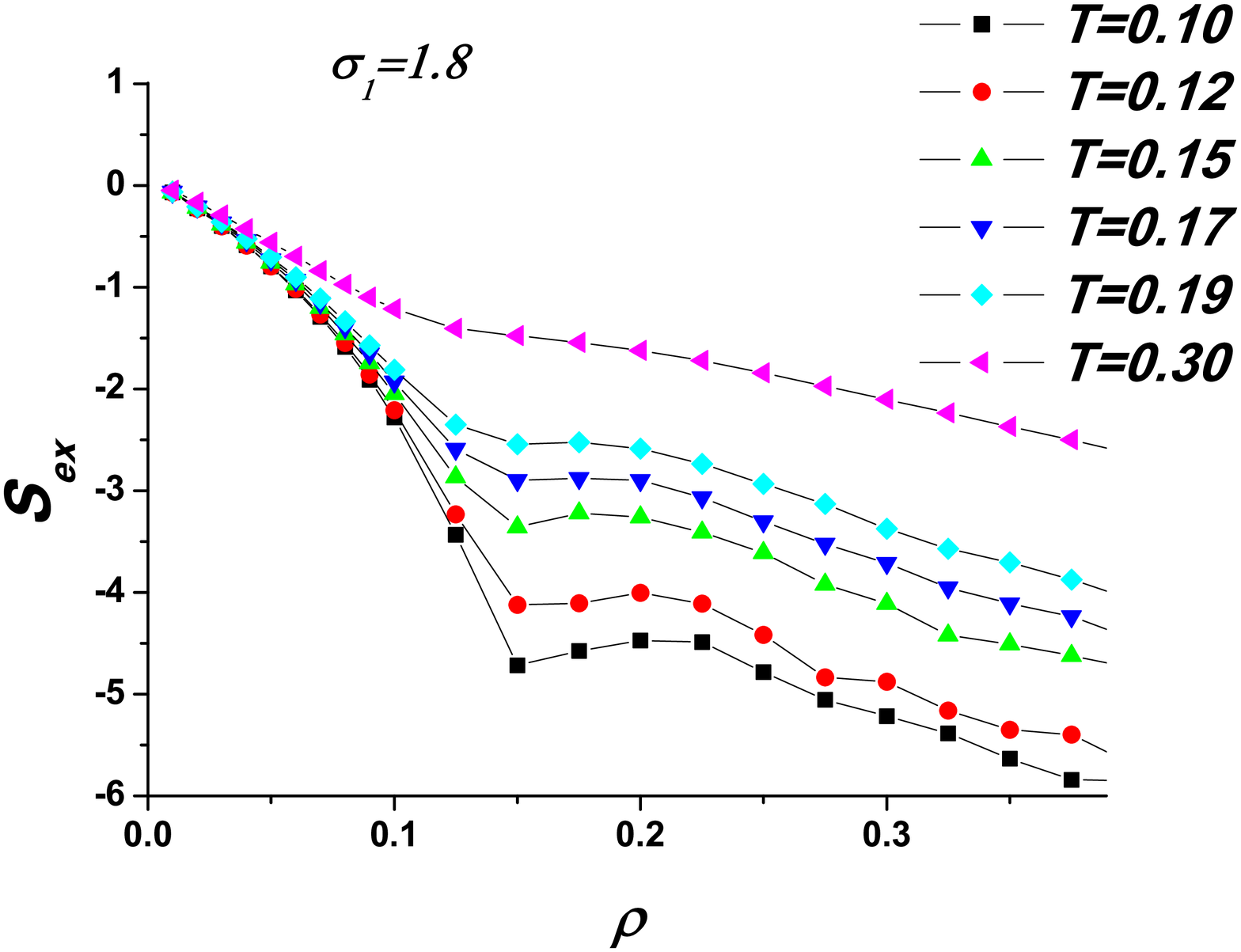}%

\caption{\label{fig:fig15} (Color online). Excess entropy along a
set of isotherms for the system with $\sigma_1=1.8$.}
\end{figure}

\begin{figure}
\includegraphics[width=8cm, height=8cm]{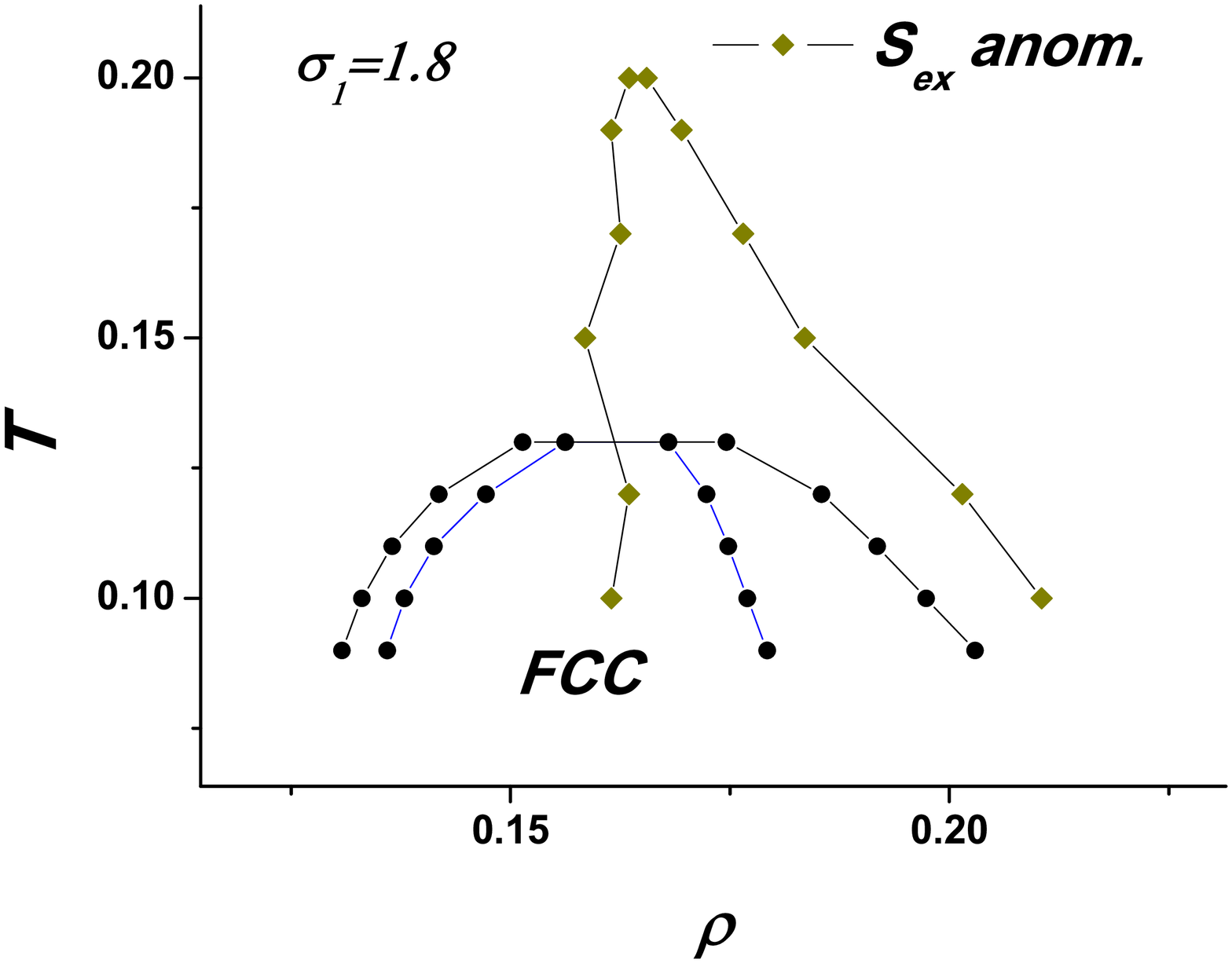}%

\caption{\label{fig:fig16} (Color online). Location of structural
anomaly region at the phase diagram of the system with
$\sigma_1=1.8$. Note that the diffusion and density anomalies are
absent for this case.}
\end{figure}

In the case of $\sigma_1=1.8$ both diffusion and density anomalies
are absent. Only excess entropy demonstrates anomalous increase
along low-temperature isotherms (Fig.~\ref{fig:fig15}). The
situation with density anomaly looks exactly the same as with the
diffusion one: the anomalous region shrinks with increasing the
step and finally hides under the melting line. At the same time
structural anomaly looks very stable with respect to the melting
line: it is still located at the temperatures almost twice higher
then the melting line maximum (Fig.~\ref{fig:fig16}).

\section{IV. Discussion}

\begin{figure}
\includegraphics[width=8cm, height=8cm]{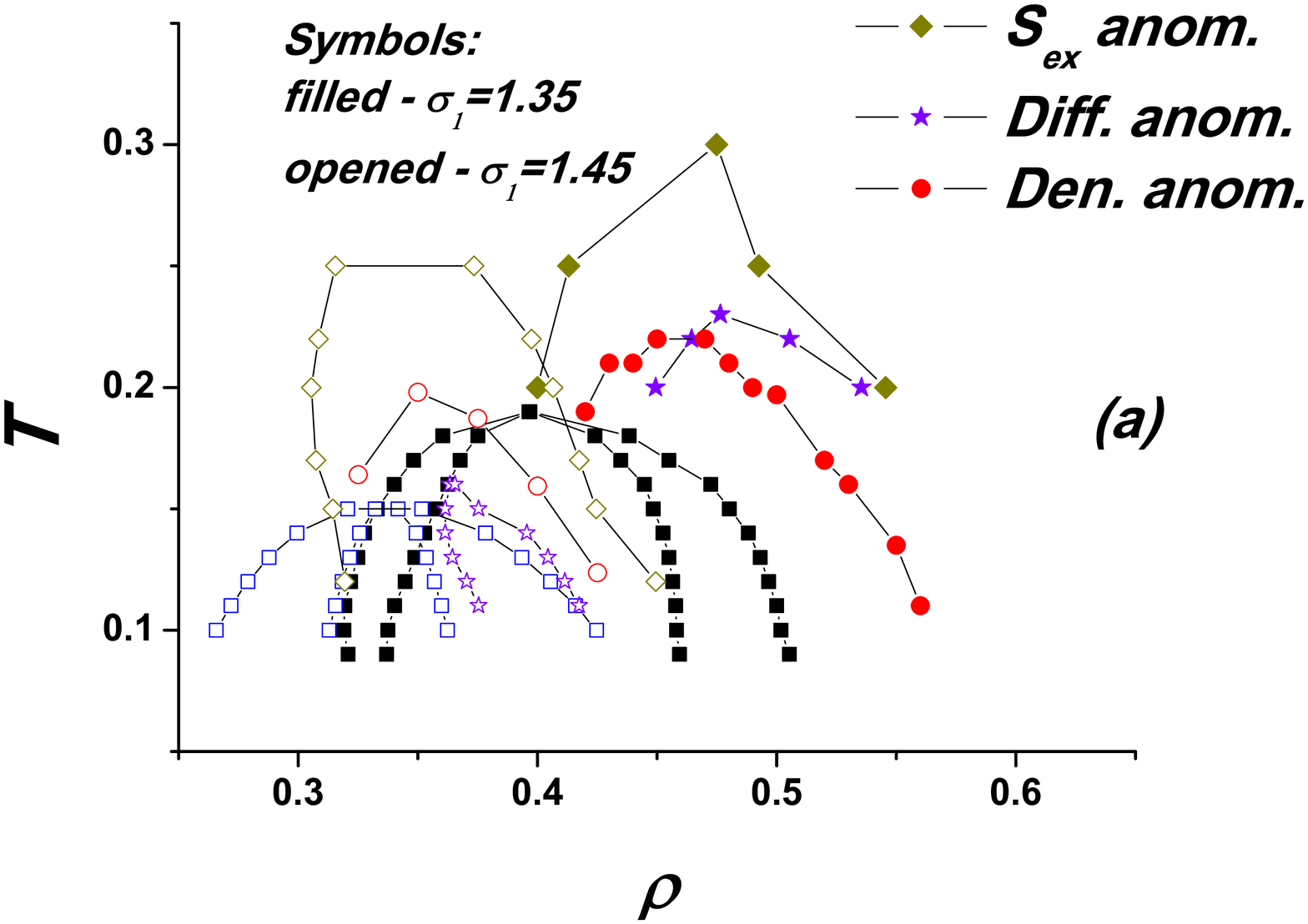}%

\includegraphics[width=8cm, height=8cm]{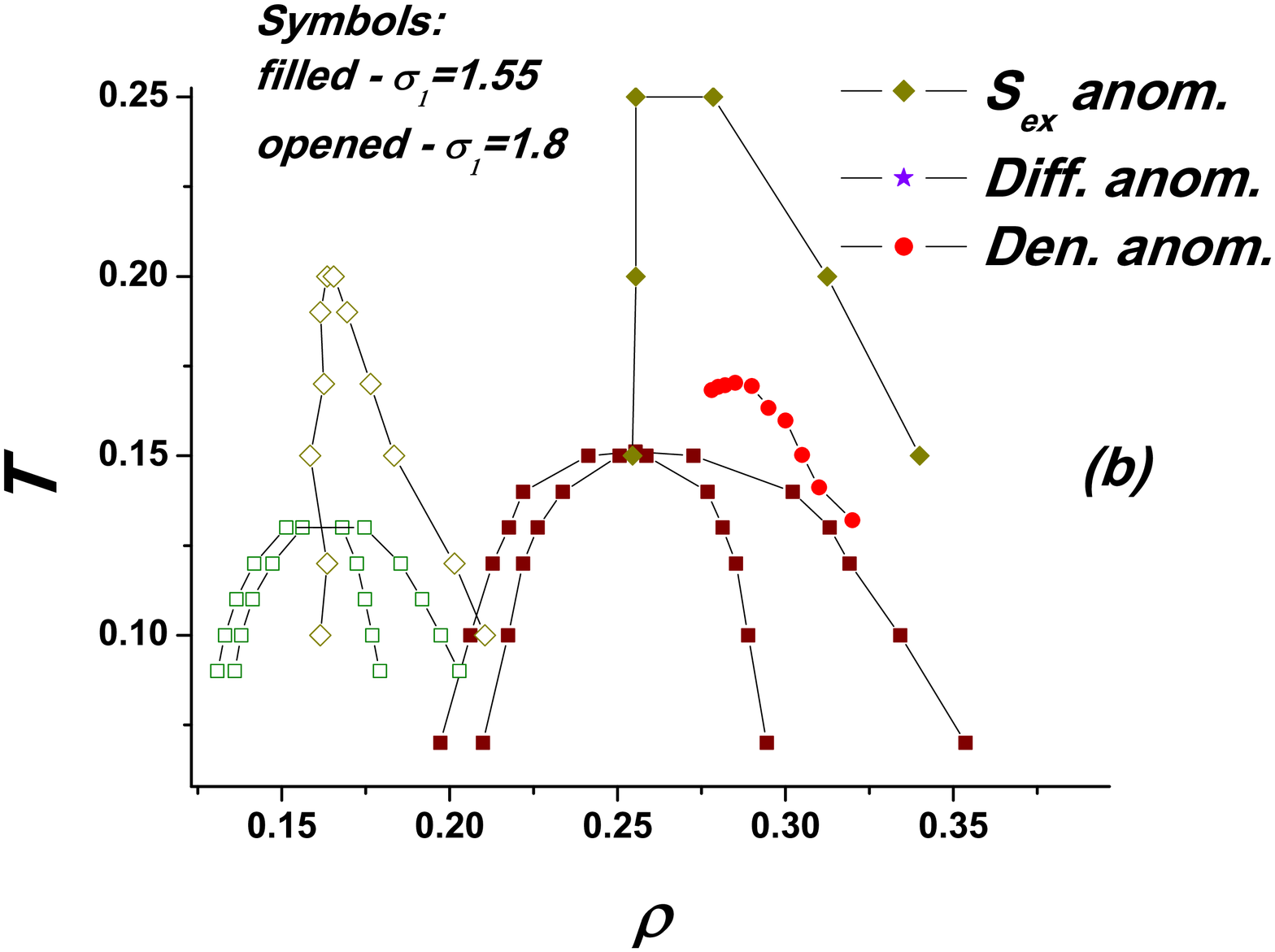}%

\caption{\label{fig:fig7} (Color online). Location of anomalous
regions at the phase diagram for (a) systems with $\sigma_1=1.35$
and $\sigma_1=1.45$; (b) $\sigma_1=1.55$ and $\sigma_1=1.8$.}
\end{figure}

Figs.~\ref{fig:fig7} (a) and (b) summarize the results for all
four systems. One can see that with increasing the step size
$\sigma_1$ the low density FCC phase and all anomalous regions
move to lower densities and lower temperatures. At the same time
the maximum temperature of both FCC phase and anomalous regions
dramatically drops from $\sigma_1=1.35$ to $\sigma_1=1.45$ and
decreases a bit from $1.45$ to $1.55$. However, the rate of
decreasing of different curves is different. The diffusion anomaly
demonstrates the fastest decay with increasing the repulsive step
size. As a result it quickly disappears under the melting line.
The density anomaly is the second fastest to decay. As result it
disappears somewhere between $\sigma_1=1.55$ and $\sigma_1=1.8$.
The structural anomaly is much more stable than the diffusion and
density ones and it does not demonstrate any precursors of
disappearance at the moment. It is interesting to note, that
similar "resisting" behavior of the structural anomaly was found
in Ref.~\cite{fr2}, where it was shown that making the soft-core
of the potential from Refs.~\cite{fr3,fr4} steeper and more
penetrable, the regions of density and diffusion anomalies
contract in the $T-\rho$ plane, while the region of structural
anomaly is weakly affected. The authors of Ref.~\cite{fr2}
concluded that a liquid can have anomalous structural behavior
without having density or diffusion anomalies. However, in
\cite{fr2} it was not found inversion of the order of anomalies
with changing the parameters of the potential. It should be also
noted that these results are consistent with results of
Ref.~\cite{barboska} where it was proposed that for some of the
core-softened systems anomalies go far under the melting line and
become unobservable.

\begin{figure}
\includegraphics[width=8cm, height=8cm]{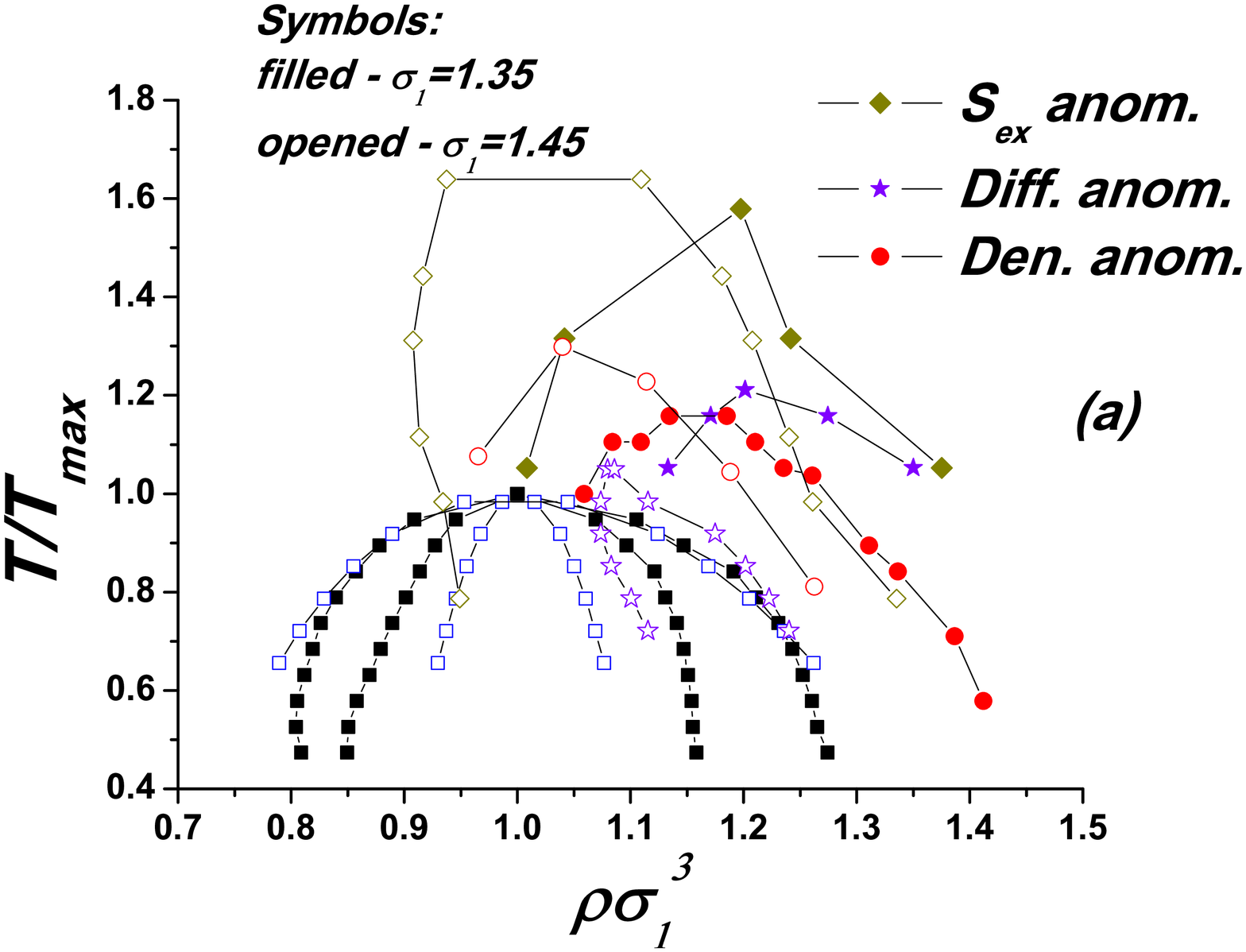}%

\includegraphics[width=8cm, height=8cm]{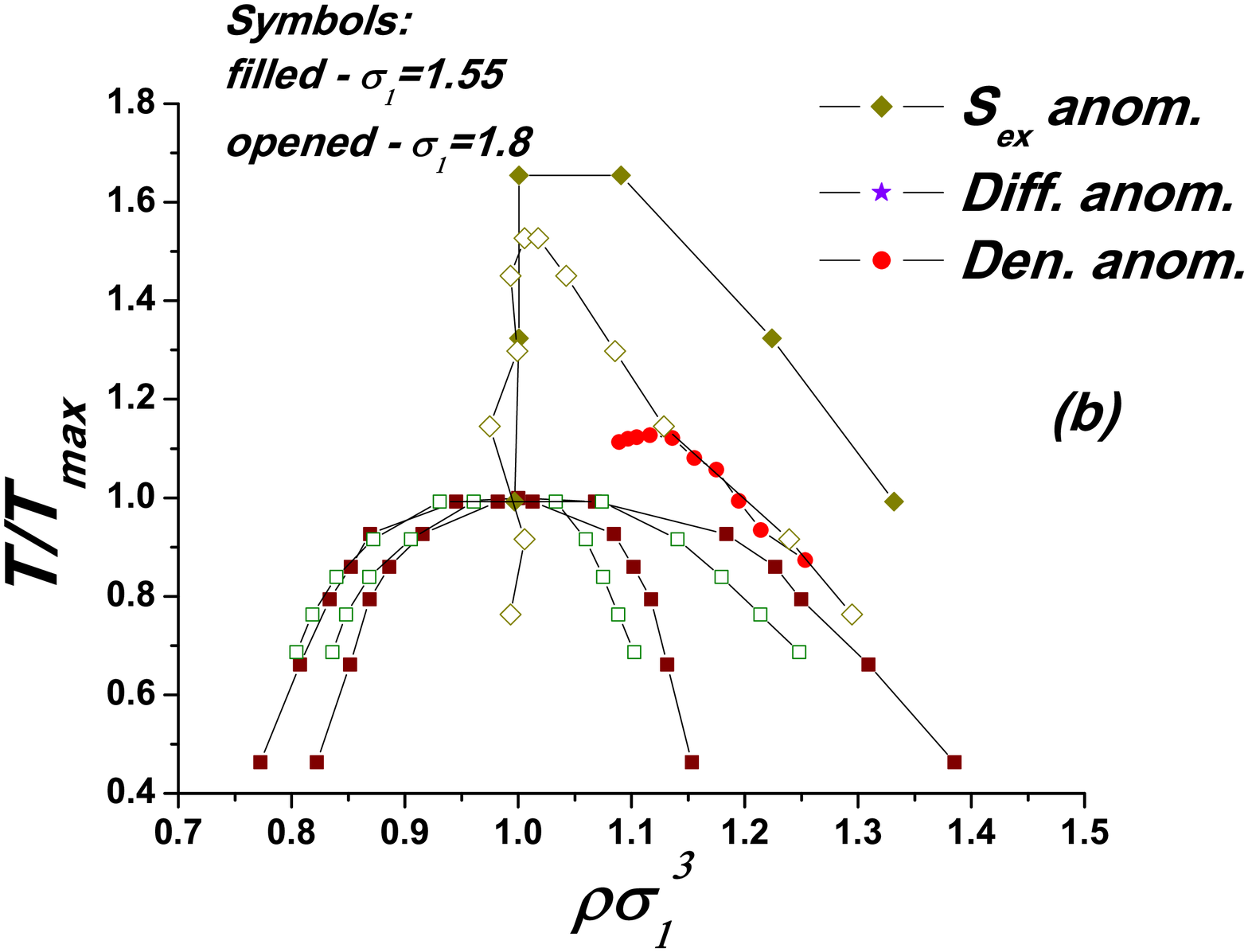}%

\caption{\label{fig:fig8} (Color online). Location of anomalous
regions at the phase diagram in scaled coordinates (see the text).
(a) $\sigma_1=1.35$ and $\sigma_1=1.45$; (b) $\sigma_1=1.55$ and
$\sigma_1=1.8$.}
\end{figure}

In our previous work we proposed to consider the anomalous regions
in scaled coordinates - $\rho \cdot \sigma_1 ^3$ and $T/T_{max}$,
where $T_{max}$ is the temperature of maximum of the melting line
of the low density FCC phase. This representation is very useful
since it allows to see the relative location of different regions
in the phase diagram. Figs.~\ref{fig:fig8} (a) and (b) give such
representation for the systems considered in the present paper.
This diagram confirms the conclusions of the previous paragraph.
Interestingly, the relative height of the structural anomaly even
increased from $\sigma_1=1.35$ till $1.55$. However, further
increase to $\sigma_1=1.8$ pushes the structural anomaly down to
the melting line. One can relate this qualitative change to the
fact that the largest step involves in the interaction the second
nearest neighbors which alters the system properties. Basing on
this change we can not generalize the results to the step of
arbitrary width: further increase of the step width will involve
more and more neighbors into interaction which can induce more
complex effects. In our previous publication we investigated
liquid-liquid transition in core-softened systems in frames of
perturbation theory \cite{wepert}. It was shown that the behavior
of the system is "periodic" with increasing the step size:
liquid-liquid transition takes place at some step size, then
disappears and on further increase appears again. We can expect
that the diffusion and density anomalies will occur again at
higher steps, however, this question requires further
investigation.

\section{V. Conclusions}

In conclusion, this publication represents a detailed study of
phase diagrams and anomalous behavior of Smooth Repulsive Shoulder
System (SRSS) which belongs to the class of core-softened systems.
We investigate the relation of the anomalous behavior with the
phase diagram and the interaction potential parameters of the
system. We show that both low density crystal phase and all
anomalous regions shrink with increasing the step size. However,
the rate of the decrease is different. It is also shown that the
order of the region of anomalous diffusion and the region of
density anomaly is inverted with increasing the width of the
repulsive shoulder.

\bigskip

\begin{acknowledgments}
We thank V. V. Brazhkin for stimulating discussions. Y.F. and E.T.
also thanks Russian Scientific Center Kurchatov Institute and
Joint Supercomputing Center of Russian Academy of Science for
computational facilities. The work was supported in part by the
Russian Foundation for Basic Research (Grants No 10-02-00694a,
10-02-00700 and 11-02-00-341a) and Russian Federal Program
02.740.11.5160.
\end{acknowledgments}

\end{document}